\documentclass[reprint,showpacs,showkeys,superscriptaddress,aps,pra]{revtex4-1}

\usepackage{graphicx}
\usepackage{epstopdf}
\usepackage{bm}
\usepackage{amssymb}
\usepackage{amsmath}
\usepackage{bbold}
\usepackage[colorlinks=true, linkcolor=blue]{hyperref}
\usepackage{ulem}
\usepackage{color, soul}
\usepackage{xcolor}
\usepackage{afterpage}
\usepackage{footmisc}
\usepackage{natbib}

\begin{document}

\title{A novel chiral spin texture: Antiferromagnetic Skyrmionium}
 \author{Mona Minakshee Manjaree Bhukta}
\author{Abhilash Mishra}
\author{Gajanan Pradhan}
\author{Sougata Mallick}
\author{Braj Bhusan Singh}
\author{Subhankar Bedanta}
\email{sbedanta@niser.ac.in}
\address{Laboratory for Nanomagnetism and Magnetic Materials, School of Physical Sciences, National Institute of Science Education and Research (NISER), HBNI, Jatni 752050, India}

\begin{abstract}
Exotic  spin textures viz. chiral domain wall, vortices, skyrmion, skyrmionium, etc. have recently emerged as active field of research because of their potential applications in high density data storage technology and logic gate computing. Magnetic skyrmionium is a skyrmion like soliton, which carries zero topological quantum number. Skyrmioniums are superior to conventional skyrmions in ferromagnets due to their negligible skyrmion hall effect and higher velocity. The physical properties of both skyrmion and skyrmionium have been investigated rigorously in ferromagnetic systems. Recent observations hint that such chiral spin structures in antiferromagnetic (AFM) systems are more promising in comparison to the ferromagnetic ones because of their robustness towards external perturbation, absence of Skyrmion hall effect, etc. However skyrmionium in AFM materials are not reported in literature so far. In this work, we demonstrate that skyrmionium can be created and stabilized in AFM materials by application of spin polarized current in an experimentally feasible geometry. We have further studied the dynamics of AFM skyrmionium by applying spin polarized current.
\end{abstract}

%\pacs{}
\maketitle
\section{Introduction }

Ever increasing demand of device miniaturization and low power consumption has led to shift in research interest from conventional storage technology towards low dimensional magnetic solitons, viz. chiral domain wall  \cite{dw1,dw2}, magnetic vortices \cite{vor1,vor2}, skrymions \cite{skyrme}, etc.  Magnetic skyrmions are twisted spin texture characterized by topological skyrmion number Q \cite{Q}, which are protected from the external perturbations. It has been observed that skyrmion propagation  in a magnetic nanotrack requires lower current density in comparison to the domain walls \cite{Q}. The topological property of magnetic skyrmion (Skyrmion number, $Q=$ 1 or -1) leads to the skyrmion hall effect (SkHE) \cite{Skhe, Skhe2}. Due to such SkHE in ferromagnetic (FM) systems, skyrmions are deflected towards the edge of the racetrack memory when driven by spin polarized current. This poses a limitation in desired logic operations. However, recent literature reports another new class of magnetic soliton; FM Skyrmionium \cite{Skyrmionium, Skyrmionium6,BogdanovJMMM1999} which are also topologically protected similar to the skyrmions. A skyrmionium is composed of 2 skyrmions: one inside another with opposite chirality and hence $Q=0$. This results in minimization of SkHE. Figure S1 (a) and (b) show the schematic representation for skyrmion and skyrmionium in FM systems. Orange and green dots represent the two distinct spin configurations aligned along $+\hat{z}$ and $-\hat{z}$ directions, respectively. Micromagnetically it has been observed that the velocity of skyrmionium is higher than that of the skyrmions \cite{Skyrmionium2}. Only a few works have been reported so far in stabilization, manipulation, and propagation of skyrmioniums in ferromagnetic systems \cite{Skyrmionium3,Skyrmionium5}.

On the other hand, the existence of antiferromagnetic (AFM) skyrmions has been predicted theoretically \cite{AFMskyrmiontheory2, AFMskyrmiontheory,BogdanovPRB2002} and studied using micromagnetic simulations \cite{AFMSkyrmion, AFMskyrmion2}. AFM skyrmions (Figure S1(c)) are another class of topological spin texture with $Q=0$ where the neighboring spins are antiparallely coupled to each other. Literature reveal that the AFM skyrmions can be propelled as a promising candidate over the FM ones due to its ultrafast dynamics, zero skyrmion hall effect, etc \cite{AFMskyrmion3, AFMskyrmion4}. On the other hand skyrmionium in AFM systems is another member in the skyrmion family (see Figure S1(d)) which is neither observed experimentally nor investigated micromagnetically. Understanding the know-how of such skyrmionium in AFMs will enrich the skyrmion physics.   

In this paper, using micromagnetic simulation we have demonstrated that skyrmionium in AFM systems can be stabilized within a range of interfacial Dzyaloshinskii-Moriya Interaction (iDMI) \cite{Dylo, Moriya} and perpendicular magnetic anisotropy (PMA). We term it as AFM-skyrmionium which is stabilized from an AFM ground state under the application of a perpendicular spin polarized current. We further show that the structure and stability of the skyrmionium strongly depends on the shape and size of the AFM nanoelement. We have also studied the skyrmionium dynamics under different current densities. 

Here we have performed the three-dimensional (3D) micromagnetic simulation by using Object-Oriented Micromagnetic Framework (OOMMF) software \cite{oommf}. It solves the time dependent Spin dynamics governed by Landau-Lifshitz-Gilbert (LLG) equation:

\begin{equation}\label{equation 1}	
\frac{d\vec{M}}{dt}=\gamma_0 \vec{H}_{eff}\times \vec{M}+\frac{\alpha}{M_s}(\vec{M}\times \frac{d\vec{M}}{dt})
\end{equation}

where $\vec{M}$ denotes the magnetization vector, $M_s$ is the saturation magnetization, $t$ is the time, $\gamma_0$ denotes the gyromagnetic ratio, $\alpha$ is the gilbert damping constant, and $\vec{H}_{eff}$ is the effective magnetic field ($\vec{H}_{eff} = \frac{-dH_{AFM}}{dM}$). $\vec{H}_{AFM}$ is the Hamiltonian of a AFM system, defined by:

\begin{flalign}\label{equation 2}
	H_{AFM}=J_{ex}\sum_{<i,j>}\vec{m}_i.\vec{m}_j +\sum_{<i,j>}D.(\vec{m}_i\times \vec{m}_j) \notag\\ - K\sum_{i}(m_i^z)^2
\end{flalign}

where $m_i=\frac{M}{M_s}$ is the normalized magnetization vector. The first term of equation (\ref{equation 2}) represents the strength of AFM interaction between two neighboring spins, where $J_{ex}$ is the AFM exchange constant ($J_{ex}>0$). The second term in the Hamiltonian represents the strength of iDMI. The last term represents the anisotropy energy associated to such system..

   \section{Methodology}

When a spin-polarized current is applied into an antiferromagnetic layer, the dynamics of the magnetization is governed by extended LLG equation (\ref{equation 3}, \ref{equation 4}), which includes an additional contribution \cite{current2,current1} of spin hall effect (SHE) and spin transfer torque (STT) to equation (\ref{equation 1}). 

When a vertical current density ($J$) is applied to the AFM layer the micromagnetism is governed by the below equation

\begin{equation}\label{equation 3}	
\frac{d\vec{M}}{dt}=\gamma_0 \vec{H}_{eff}\times \vec{M}+\frac{\alpha}{M_s}(\vec{M}\times \frac{d\vec{M}}{dt})+\frac{u}{aM_s}(\vec{M}\times \hat{p} \times \vec{M})
\end{equation}

where, $u(=\frac{J\theta_{SH}{\mu}_{B} g}{2eM_s})$ lies along the direction of electron ($e$) motion, $\hat{p}$ stands for the polarization direction of the current, $\theta_{SH}$ is the spin Hall angle, and $g$ is the Lande' g-factor. In our work we have taken $\hat{p} =+\hat{z}$ for the nucleation of AFM-skyrmionium.  After the nucleation the skyrmionium can be propagated in the nanotrack by applying a current either in out-of-plane or in-plane direction. If one defines the current density for the skyrmionium motion to be $J_m$, then two cases may be considered: (Case I) $J_m = +\hat z$ for ; (Case II) $J_m = +\hat x$.  For case I we considered the $\hat{p}$ to be along the $\hat y$ direction. Then the micromagnetism can be explained by equation 3. However for case II where the $J_m =+\hat x$, the micromagnetism of the skyrmionium can be explained by the following equation \cite{fert2017}.

\begin{flalign}\label{equation 4}
		\frac{d\vec{M}}{dt}=\gamma_0 &\vec{H}_{eff}\times \vec{M}+\frac{\alpha}{M_s}(\vec{M}\times \frac{d\vec{M}}{dt})\notag\\&+\frac{u}{M_s^2}(\vec{M}\times \frac{\partial \vec{M}}{\partial x} \times \vec{M})-\frac{\beta u}{M_s}(\vec{M} \times \frac{\partial \vec{M}}{\partial x})
\end{flalign}

where, $\beta$ is the Slonczewski-like STT coefficient denoting the strength of the non-adiabatic STT.

The micromagnetic simulations presented in this work have been performed using a set of the OOMMF extensible solver (OXS) in the 3D layout of OOMMF \cite{oommf}. The simulations are performed on a 0.5 $nm$ thick $KMnF_3$ G-type antiferromagnetic thin film of area 100 $nm\times$ 100 $ nm$. IDMI \cite{DMI} and spin polarized currents \cite{anv} have been incorporated using different oxs extension modules of OOMMF. The parameters considered in the simulations are obtained from reference \cite{material}: $M_s=0.376\times 10^6$ $Am^{-1}$, exchange stiffness constant $A=-6.59\times 10^{-12}$ $Jm^{-1}$, $\theta_{SH}=0.07$. The PMA constant $(K)$ and iDMI constant $(D)$ have been varied in the range of (0.03-0.33)$\times 10^{6}$ $Jm^{-3}$ and (0.9-2.1) $mJm^{-2}$, respectively. It should be noted that, the simulations presented here do not include any periodic boundary condition (PBC).

\section{Results and discussion}
\begin{figure*}
	\centering
	\includegraphics[width=0.9\textwidth]{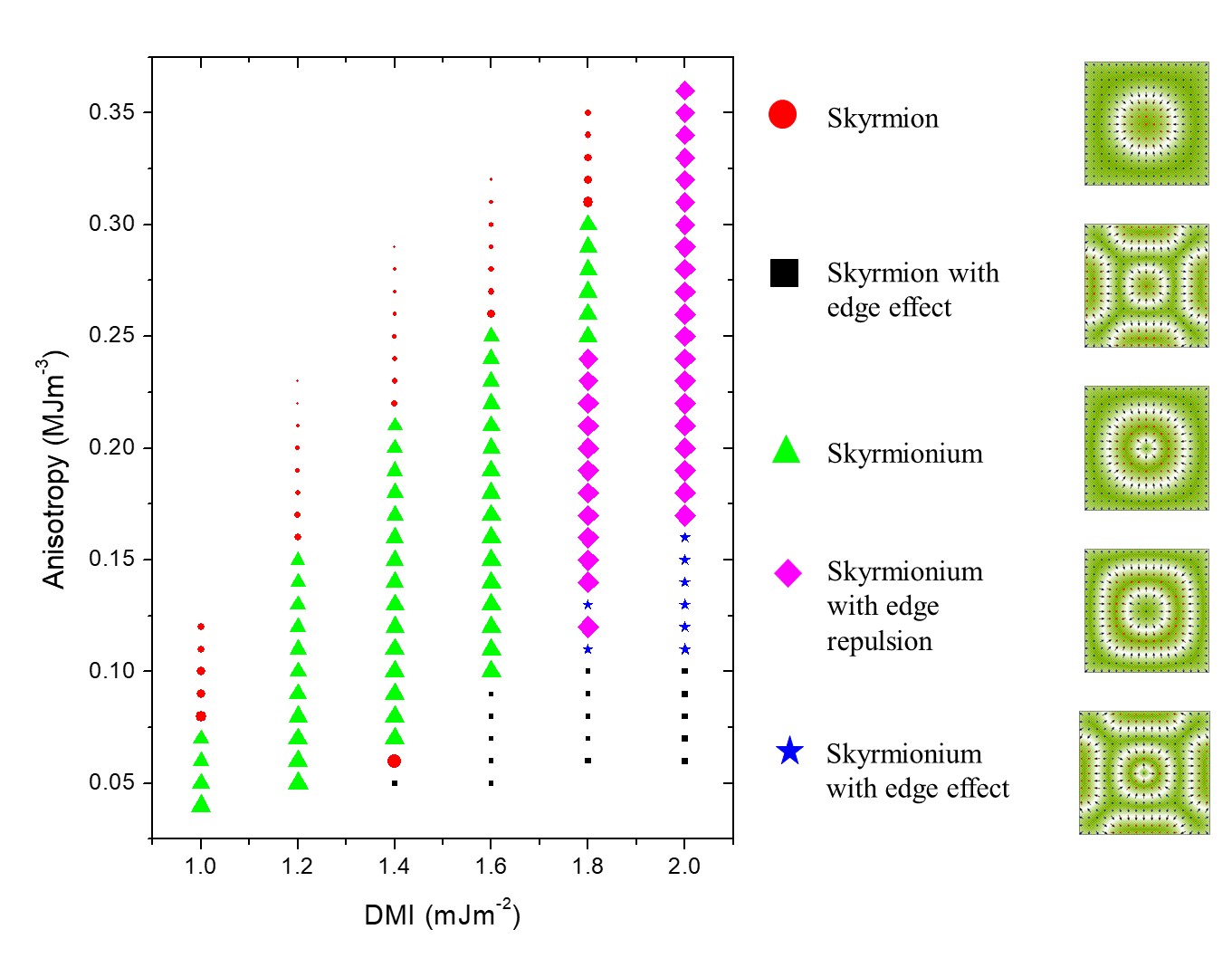}
	\caption{Phase diagram showing the ground state spin structures in AFM systems for a range of D and K. Various spin textures i.e. skyrmion, skyrmion with edge effects, skyrmionium, skyrmionium with edge effect, and skyrmionium with edge repulsion states are represented by red circle, black square, green triangle, magenta diamond, and blue star shape, respectively, The radius of the spin textures are normalized and corresponds to the size of individual structures. The blank data points represent the ground state of the antiferromagnets.}
	\label{fig1}
\end{figure*}

An AFM ground state (see figure S2(a)) consist of two sub-lattices where the spins are aligned antiparallel to each other. To nucleate the skyrmionium we have considered a square element having the length of  100 $nm$ with AFM ground state. Then a current density of $J=40\times10^{12} Am^{-2}$ in case I geometry  is applied for 3.48 $ps$ in a disc shaped region of diameter $d$=25 $nm$ at the centre of the square nanoelement (see supplementary figure S2(b)).  As soon as the current is applied, the spins start flipping at the injected region (see figure S2(c)) to minimize the energy of the system for the total duration of the pulse. Once the current pulse is stopped, the spin configuration starts relaxing (see figure S2(d)) and reaches a stable ground state configuration (Figure S2(e)). We observe that this stable ground state is a novel chiral spin structure which is similar to the Neel skyrmionium \cite{Skyrmionium} reported for FM systems. Hence we call this new spin configuration as AFM-skyrmionium.

The stability of a skyrmionium in an AFM ground state has been investigated in a 100 $nm$  $\times$ 100 $nm$ square sample of thickness of 0.5 $nm$. Figure \ref{fig1} shows the phase diagram of different magnetic states (skyrmion, skyrmionium, skyrmionium with edge effect, skyrmionium with edge repulsion and skyrmion with edge effect) evolved from the AFM ground state under the applied current density of $J=40\times10^{12}Am^{-2}$ for  3.48 $ps$. Here the strength of $D$ and $K$ have been varied over a wide experimentally feasible range of $(0.9-2.1) mJm^{-2}$ and $(0.03-0.33)\times10^6Jm^{-3}$, respectively.  In figure \ref{fig1} we define the skyrmion, skyrmion with edge effects, skyrmionium, skyrmionium with edge effect and skyrmionium with edge repulsion states by red circle, black square, green triangle, magenta diamond , and blue star shape, respectively. The structure of the different magnetization states are also depicted in this figure.

It is understood that $D$ helps in stabilizing the skyrmionium as well also determines the size of it. Further we observe that for obtaining a stable AFM skyrmionium the lower bound of $D$ value is 1.0 $mJm^{-2}$ . Simulations reveal that, for higher value of $D$ in the range of 1$<D<$1.6 $mJm^{-2}$, skyrmionium can be stabilized in a larger range of $K$. We observe that stable skyrmionium configurations occur for a narrower range of $D$ in comparison to that of the skyrmions. However for $D>$1.8 $mJm^{-2}$ , the edge/finite size effects dominate and the shape of the skyrmionium gets distorted. Under such edge effect, we obtained two different structure of the skyrmioniums defined as skyrmionium with edge effect and edge repulsion states. We have further simulated the ground states by considering even higher values of D $(3<D<8)$ $mJm^{-2}$ and $K$ ($0.1 < K <0.8$ )$\times10^{6}Jm^{-3}$ to understand their role in tuning the exact features of the edge effects. The ground states of such structures are shown in supplementary figure S3. We note that the feature size of the spin textures decreases forming maze like magnetic configurations with enhancement of $D$.

\begin{figure}[h!]
	\centering
	\includegraphics[width=0.45\textwidth]{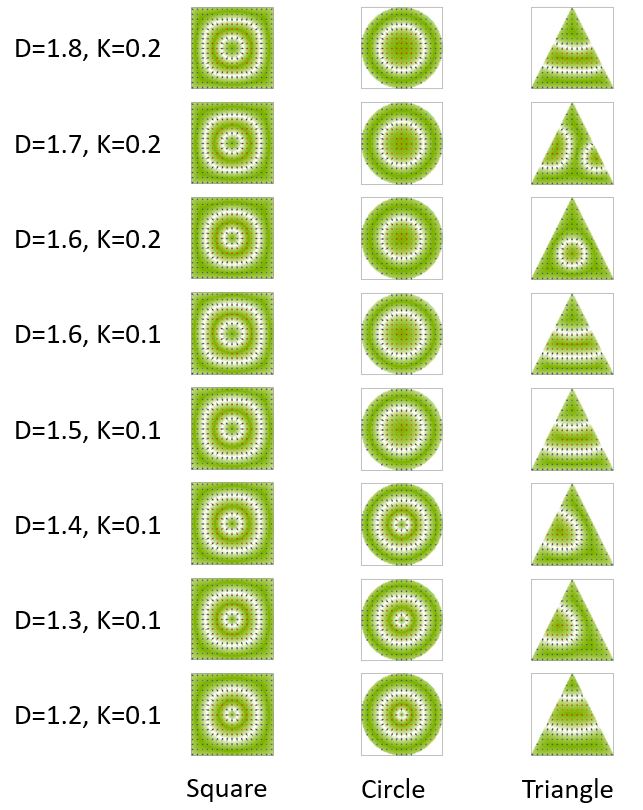}
	\caption{Ground state spin configurations at various $D$ ($mJm^{-2}$) and $K$ ( $\times 10^6Jm^{-3}$) for square, circular, and triangular shaped AFM nanoelements.}
	\label{fig2}
\end{figure}

Further, it is known that the shape of the magnetic nanoelement/nanotrack plays an important role in determining the final magnetic configurations of such spin textures \cite{aroop}. To understand the effect of the shape on the AFM skyrmionium, we have simulated the ground states for a few other trivial structures viz. circle and triangle. To elucidate the complete spectrum, we have also varied the values of $D$ and $K$ in the range where stable skyrmionium ground states are observed for the square shaped elements. Figure \ref{fig2} shows the comparison between the ground states in square, circular, and triangular shaped nanoelements for various combinations of $D$ and $K$. We observe that the formation of skyrmionium occurs only for $D$ = (1.2-1.4) $mJm^{-2}$ and $K$ =0.1 $\times 10^6 Jm^{-3}$ in case of the circular nanoelements. Other combinations of $D$ and $K$ lead to occurrence of skyrmion rather than the skyrmionium. Additionally, we did not obtain any skyrmionium in the triangular nanoelements for various values of $D$ and $K$ considered in this paper. Hence, it can be concluded that in addition to the combination of $D$ and $K$, the shape of the nanoelements also plays a pivotal role in stabilization of the AFM-skyrmionium.
\begin{figure}[h!]
	\centering
	\includegraphics[width=0.45\textwidth]{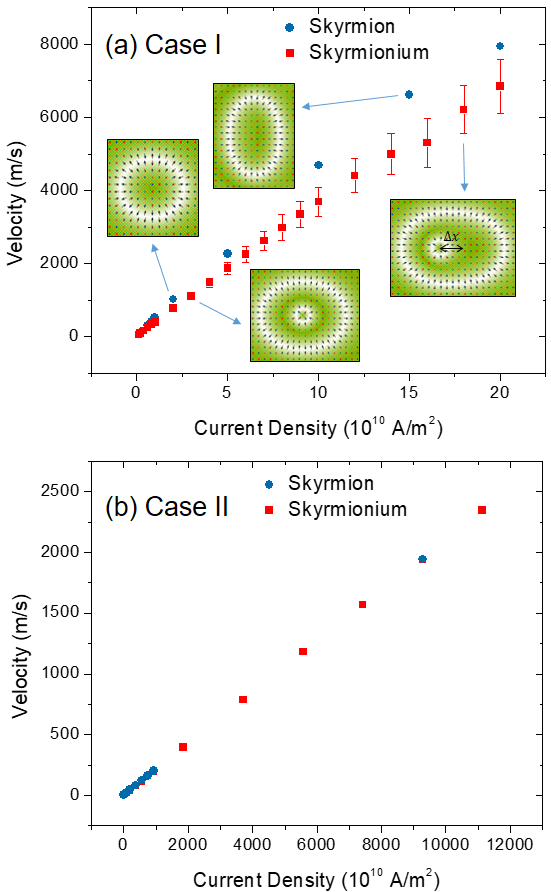}
	\caption{Velocity of the AFM skyrmion (blue circles) and AFM-skyrmionium (red squares) at $D=1.08$ $mJ/m^2$ and $K=0.116$ $\times10^6 J/m^3$ where the spin polarized currents are applied in (a) case I and (b) case II. Shape of both skyrmion and skyrmionium at different current densities are shown in (a). The red vertical lines represent the error bar in velocity which arises due to the distortion ($\Delta x$) in the shape of the skyrmionium.}
	\label{fig3}
\end{figure}    

The nucleation of AFM-skyrmioniums has been achieved by applying a spin polarized current in the $+ \hat{Z}$ direction. However, to study the dynamics of such skyrmionium we have applied a spin polarized current along $+ \hat{Z}$ (Case I) and $+ \hat{X}$ (Case II) directions. For this study a rectangular nanoelement (similar to nanowire) having a length of 300 $nm$ and width of 100 $nm$ has been considered. Here we have taken the values of $D$, $K$ and $\beta$ to be 1.08 $mJ m^{-2}$,  0.116 $\times 10^6 J m^{-3}$, and  0.02, respectively. We have also made simulations keeping the same values of $D$, $K$, and $\beta$ to nucleate AFM skyrmions by applying $J$ for a little more duration (4.2 $ps$) and the diameter of the contact pad was 20 $nm$. We also studied the AFM skyrmion dynamics to compare with the AFM-skyrmionium behavior.

Figure \ref{fig3} (a) and (b) shows the velocity for both the AFM skyrmion (blue circles) and skyrmionium (red squares) where the spin polarized current is applied in the configuration defined by case I and II, respectively. The velocity is calculated by measuring the displacement of the centre of the skyrmion/skyrmionium in $+X$ direction with respect to the duration of the applied spin polarized current. In case I the velocity of the skyrmion and skyrmionium remains comparable at the low current density regime ($<2\times10^{10}$ $Am^{-2}$). However, the velocity of the skyrmionium decreases as the current density is further enhanced. It should be noted that this behavior is in contrary to what has been observed for the FM systems where the velocity of the skyrmionium is higher than that of the skyrmion \cite{Skyrmionium2}. The skHE in the FM skyrmionium is significantly lower than that of the skyrmion. However, in the AFM case the skHE for both skyrmion and skyrmionium is zero. Nevertheless in AFM-skyrmionium the higher magnetostatic energy hinders the fast movement. This may be the possible reason behind the occurrence of such velocity behavior in the AFM case. The red vertical lines represent the error bar in velocity which arises due to the distortion ($\Delta x$) in the shape of the skyrmionium. 

\begin{figure}
	\centering
	\includegraphics[width=0.45\textwidth]{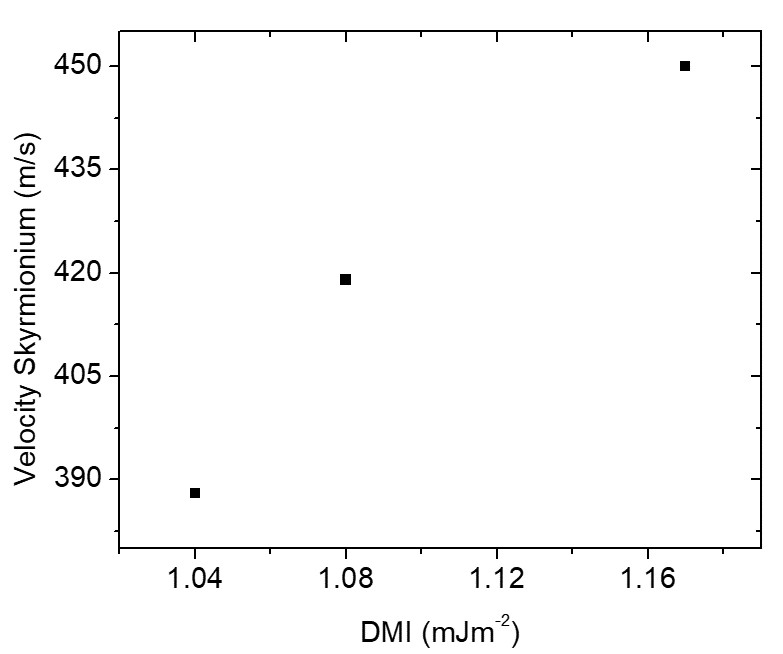}
	\caption{Velocity of the AFM skyrmionium as a function of $D$ with constant current density of $1\times 10^{10} Am^{-2}$ in case I}
	\label{fig4}
\end{figure}

On the other hand the velocities of both skyrmion and skyrmionium in AFM are exactly same in the whole range of applied current for case II (figure \ref{fig3}(b)). Comparison of the velocity plot for both the cases reveals that low threshold current density is required for skyrmionium propagation in the geometry defined by case I. Therefore the case I geometry is more power efficient for device applications. 

Further, we have investigated the velocity of the skyrmionium as a function of $D$ with a fixed current density of $1\times10^{10} Am^{-2}$ for case I geometry. The velocity as a function of $D$ is plotted in figure \ref{fig4}. It is observed that the velocity increases from $388$ $m/s$ to $450$ $m/s$ with increasing $D$ from $1.04-1.17$ $mJm^{-2}$.

\section{Conclusion}

For the first time, we have shown the possibility of stabilizing skyrmionium in antiferromagnetic systems by the help of micromagnetic simulations. It has been observed that the stability of such novel spin texture strongly depends on $D$, $K$, and shape of the magnetic nanoelements. Extensive simulations confirm that there is a narrow range of values of $D$ and $K$ for which stable skyrmionium ground states arise. Propagation of the AFM skyrmioniums in rectangular nanotrack has been achieved by applying spin polarized currents. We also show that the velocities of the skyrmionium and skyrmion are comparable to each other Further, we confirm that the threshold current density required to drive the skyrmioniums is less for case I compared to case II. The phase map for various magnetic configurations and their dynamics presented in this paper can be utilized as a guide map to design experimental systems for skyrmion as well as skyrmionium based future spintronics and logic applications.

\section*{acknowledgments}
The authors thank DAE, Govt. of India and the Indo-French collaborative project supported by CEFIPRA for providing the research funding. The authors also thank Mr. Aroop Kr. Behera for his help in the simulations.

\end{document}